# Is MOOC Learning Different for Dropouts? A visually-driven, multi-granularity explanatory ML approach

Ahmed Alamri, Zhongtian Sun, Alexandra I. Cristea, GauthamSenthilnathan, Lei Shi and Craig Stewart

Department of Computer Science Durham University
Corresponding author: ahmed.s.alamri@durham.ac.uk

**Abstract.** Millions of people have enrolled and enrol (especially in the Covid-19 pandemic world) in MOOCs. However, the retention rate of learners is notoriously low. The majority of the research work on this issue focuses on predicting the dropout rate, but very few use *explainable learning patterns* as part of this analysis. However, *visual representation of learning patterns* could provide deeper insights into learners' behaviour across different courses, whilst numerical analyses can – and arguably, should – be used to confirm the latter. Thus, this paper proposes and compares *different granularity visualisations for learning patterns* (based on clickstream data) for both course *completers* and *non-completers*. In the large-scale MOOCs we analysed, across various domains, our *fine-grained, fish-eye visualisation* approach showed that non-completers are more likely to jump forward in their learning sessions, often on a *'catch-up' path*, whilst completers exhibit *linear behaviour*. For *coarser, bird-eye granularity visualisation*, we observed learners' transition between types of learning activity, obtaining *typed transition graphs*. The results, backed up by statistical significance analysis and machine learning, provide insights for course instructors to maintain engagement of learners by adapting the course design to not just 'dry' predicted values, but explainable, visually viable paths extracted.

**Keywords:** Learning Analytics, Visualisation, MOOCs, Behavioural Pattern, Machine Learning

## 1    Introduction

Massive Open Online Courses (MOOC) platforms (e.g., edX, Coursera, FutureLearn, Udemy) offer vast amounts of virtual course materials. With easy access, especially during the Covid-19 pandemic, MOOCs are the *de facto* platform for self-learning. However, in spite millions of enrolments, the completion rate is usually less than 10%, [1]. Researchers have attempted early identification of likely dropouts, to allow for interventional activities of course instructors. Existing studies [2, 3, 4, 5, and 6] mainly employ machine learning and statistical analysis. However, few can provide sound insights and explanations to course instructors regarding the learning manner of participants [7]. Importantly, to the best of our knowledge, there is no prior work on *visually* comparing the learning path of completers and non-completers for the entire course session. This paper addresses the following research questions:

*1. How can learning paths be visualised to differentiate between completers' and non-completers' behaviour (to inform teachers for early interventions)?*
*2. Are there (significant) differences in the learning paths of completers and non-completers and can they be deduced from visualisation early on in the course?*
*3. What kind of level of granularity is necessary for the visualisation of significant differences between completers and non-completers?*

The main contributions of this paper are:

- Providing new insights into *early learning behaviour* exhibited by course *completers* and *non-completers* through *bird-eye* and *fish-eye visualisation* of *partial* or *full learning graphs*, with different levels of information disclosure.

- Proposing *visual graph analysis as a pre-step to machine learning and prediction*, here illustrated by discovering *linear* or *catch-up* behaviours, which then can be reliable predicted.

- Showing that *theme*-based visualisation (which also can be at bird-eye or fish-eye level) detects other relations in the course, e.g. the effect of forums.

## 2 Related work

Existing studies on MOOCs analytics mostly focus on finding reliable completion indicators from learner behaviour patterns, using data of forum activities[8], click-streams [9], participants' time spent and number of accesses [10], assignment activities [11] to name a few. Other predictive studies [12, 13, 14] attempt to forecast the performance, including final grade and pass/fail in exams. Overall, existing research usually does not disregard the potential in using visualisation as an initial step before prediction, nor do they consider the granularity of visualisation as a factor.

### 2.1 Visualisation

The explanatory power of visualisation has been stated to be crucial to provide more insights for module instructors and designers, next to completion prediction [7]. A learning path is defined as the learning trajectory through a course by learners; according to [15], participants generally study web courses in a non-linear manner. We are specifically interested in comparing the transition behaviour between completers and non-completers, as suggested by [16]. [17] investigated the typical learning activity sequences across two MOOC datasets and mined the difference in learning themes among groups with different grades, by visualising the theme distribution. However, they mainly focused on which topics were more popular, instead of visualising the entire learning paths of their four groups of learners (none, fail, pass and distinction). Later, [7] visualised log traces of learners across four edX MOOC datasets, using discrete-time Markov Models and observed that learners were more likely to jump forward than backwards from video content. Additionally, they found that non-passing learners were more likely to exhibit binge video watching, i.e., transit from one video to another without answering questions, deviating from the designed linear learning path. However, they only visualised the video interaction activity of passing and non-passing learners, instead of the whole learning sessions, like in our work. Recently, [18] used visualisation software, Gephi, to visualise clickstream-based learning paths. They observed that learners are more likely to skip the quizzes at the end of each chapter, by watching the beginning videos of the next chapter, but learn linearly within chapters. However, neither did such previous studies explore the phenomenon in-depth, nor provided a convincing explanation. This paper validates that completers behave differently from dropout learners, by visualising the entire learning paths of participants. We also implement machine learning models, to statistically analyse students' movements, by combining courses' themes and content.

### 2.2  Statistical Analysis

According to [19], statistical analysis can be divided into descriptive statistics, to summarise the demographics information of learners and inference statistics, to explore behaviours exhibited by participants. For instance, [20] firstly examined if there is any impact on the behaviour of learners, after they reached the passing state. They examine weekly quiz score distribution for all learners by K-means clustering, which showed that early passers obtained relatively lower scores immediately after passing, compared with their previous performance. Later, [21] investigated the motivation of two groups of completers: university students and general participants, using Mann–Whitney U test, and concluded that as participants' ages increased, earning a certificate was less significant. Recently, [22] explored different behaviours of course completers and non-completers from a discourse perspective on course content review, via a chi-square test. They found that completers were more likely to post original and less negative opinions, whilst non-completers were more willing to reply to and criticise others' posts. Inspired by their work, we implement a Wilcoxon signed-rank test for completers and non–completers to learn their transition behaviours among different learning themes, separately. Additionally, we also apply two machine learning algorithms for early completion prediction, based on our analysis of two groups of student activities during the first week.

## 3  Methodology

To explore if completers and non-completers behave differently, we apply visualisation analytics firstly, to identify the different learning paths executed by these groups and then implements statistical modelling to analyse their learning behaviour (transition from themes such as *video, quiz, discussion, review and article*) across different courses. Additionally, by comparing dropout learners' transition from different learning activities across different courses, the paper offers insights into the impact of the designed course learning path on the learning behaviour of participants.

### 3.1 Dataset

The dataset of learner activities has been extracted from 8 runs of 4 Future Learn-delivered courses from The University of Warwick between 2013 and 2017, with over than 106,036 learners. Runs are defined as the number of repetitive teaching session for each course. We investigate learner activities across different domains: Psychology (The Mind is Flat and Babies in Mind), Literature (Shakespeare and His World) and Computer Science (Big Data). The over 1.5 million learner activities include watching videos, taking a quiz, discussion, submitting assignments, viewing assignment feedback and reviewing another learner's assignment and giving feedback.

**Table 1.** The dataset of learner activities

| Course Title & Run | | Enrolled | Accessed 1 step | Dropout | Completers |
|---|---|---|---|---|---|
| Babies in Mind | Run 1 | 12651 | 5841 | 4634 | 1207 |
| | Run 2 | 9740 | 4924 | 4030 | 894 |
| Big data | Run 1 | 11281 | 4715 | 4202 | 513 |
| | Run 2 | 5761 | 3840 | 3583 | 257 |
| Shakespeare and His World | Run 1 | 15914 | 9050 | 7170 | 1880 |
| | Run 2 | 12692 | 6902 | 5804 | 1098 |
| The Mind is Flat | Run 1 | 22929 | 8198 | 6858 | 1340 |
| | Run 2 | 15068 | 6760 | 5743 | 1017 |

Of all participants, 50230 learners have accessed at least one step of the course, but over half, 53%, have never accessed the course after registration. The 42024 learners who accessed less than 80% are defined as *dropout learners*. The 8206 learners who accessed 80% or more of the materials in one run are defined as *completers*. The 80% threshold (as opposed to, e.g., 100% completion) is based on prior literature discussing different ways of computing completion [23]. Completers represent 7% of the total and 16% of the learners who accessed at least once.

### 3.2 Visualisation of High & Low Granularity Levels

After analysing the learning path of learners, we have divided the dataset of learners' activities into two components: *linear* and *catch-up* (Figure 1). The former shows the linear path between two sequential steps and the latter shows the catch-up activities, i.e., jumping behaviour. The paper implements a flow network analysis, to present the learning pattern for linear (source: $x_{i-1}$, destination: $x_i$) and catch-up (source: $x_{i-1}$, destination $\neq x_i$). Depending on the granularity, the learning pattern is further identifiable as *bird's eye view*, i.e., high granularity (a node representing multiple steps), and *fisheye view*, i.e., low granularity view (step-level representations). In addition, course activities have been grouped by different colours, based on their themes.

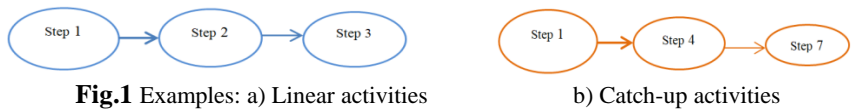

**Fig.1** Examples: a) Linear activities          b) Catch-up activities

The size of the circle represents the number of learners who accessed the course content and the thickness of the arrow shows the percentage of learners' movements.

### 3.3 Statistical Analysis and Machine Learning

We applied the Shapiro–Wilk test for normality check. For non-parametric data, we applied the Wilcoxon signed-rank test for significance measurement of differences between groups. We report results as percentages, instead of a total number of learners, to eliminate the effects of having different numbers of learners for each course. To explore more on learners' engagement with different learning themes, we also build an early dropout prediction model based on the time spent on each activity, by *ensemble machine learning methods*: Gradient Boosting [24] and XGBoost [25].

## 4 Results and Discussion

The flow network analysis shows that completers are more likely to complete the courses *linearly* (Figures 4, 5), whilst non-completers are more likely to skip quizzes and assessments (the *catch-up* learning pattern; Figures 2,3 which mainly show the overall learning pattern instead of providing a clear view in details due to the course length). For instance, non-completers have various learning paths; some of them may directly jump to lessons in week two after accessing the first lesson (Figure 3). Instead, completers are much more "obedient", as they mainly follow the designed learning path, compared with dropout learners; interestingly, this holds true across different domains - as the bird eye views in Figure 2 show. Statistical analysis results in Table 2 further confirm that these learning paths are significantly different.

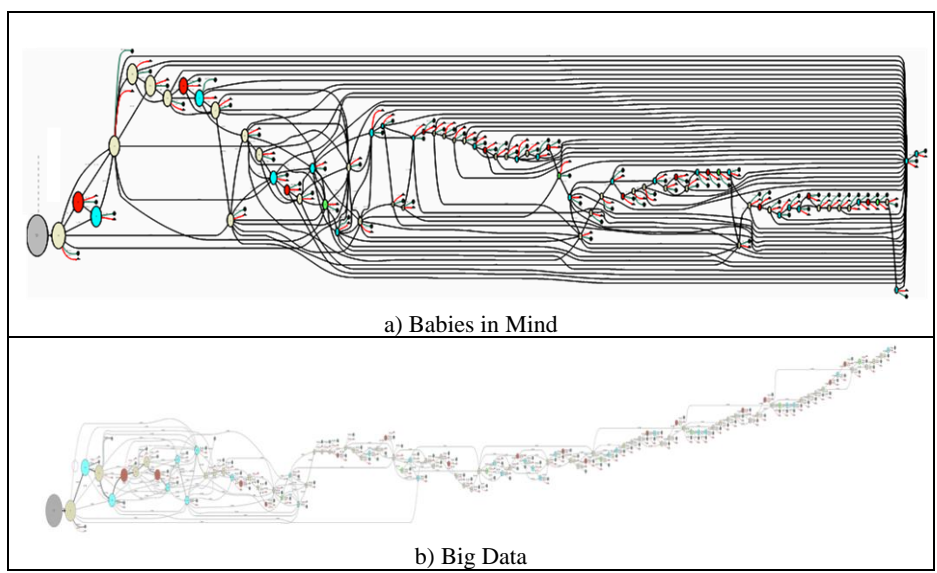

a) Babies in Mind

b) Big Data

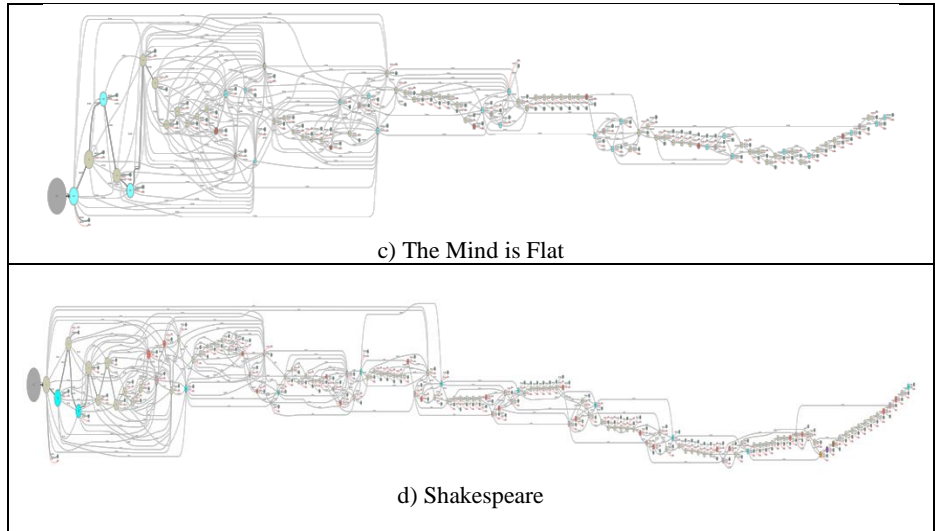

c) The Mind is Flat

d) Shakespeare

**Fig.2.** Dropout learners learning path (bird eye view)

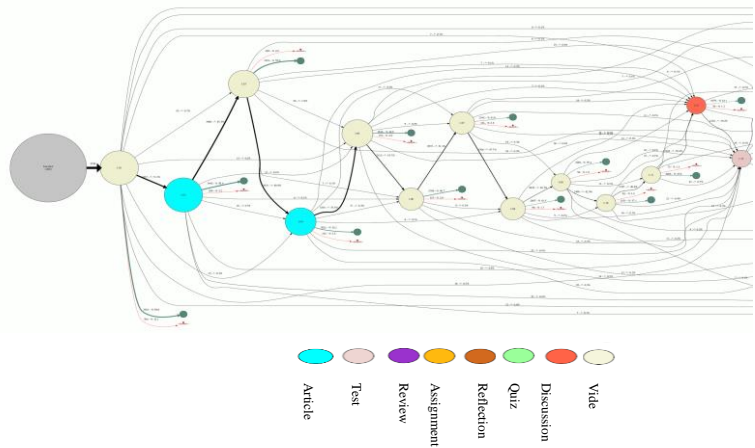

**Fig.3.** Dropout learning path, first week (Shakespeare and his world) fish eye view

**Table 2.** P-values of linear and catch-up learning activities

| P-value | Catch-Up activities | Linear activities |
|---|---|---|
| Babies in Mind Run 1 | 1.13E-13 (p<0.001) | 2.46E-85 (p<0.001) |
| Babies in Mind Run 2 | 7.74E-14 (p<0.001) | 2.32E-62 (p<0.001) |
| Big Data Run 1 | 2.66E-018 (p<0.001) | 5.97E-110 (p<0.001) |
| Big Data Run 2 | 1.35E-68 (p<0.001) | 2.66E-18 (p<0.001) |
| Shakespeare Run 1 | 1.130E-13 (p<0.001) | 2.09E-23 (p<0.001) |
| Shakespeare Run 2 | 7.73E-14 (p<0.001) | 1.87E-09 (p<0.001) |
| Mind is Flat Run 1 | 2.66E-018 (p<0.001) | 5.21E-74 (p<0.001) |
| Mind is Flat Run 2 | 6.62E-63 (p<0.001) | 2.51E-16 (p<0.001) |

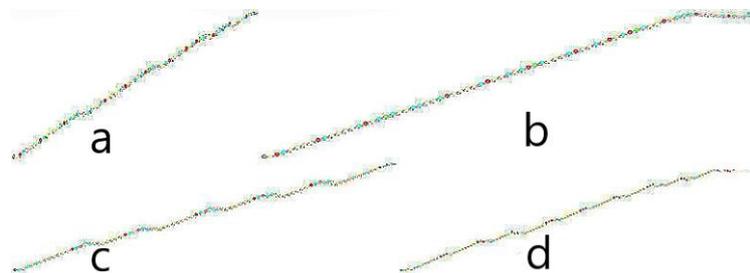

**Fig.4.** Completers learners learning path (Bird eye view)
a: Babies in Mind   b: Big Data   c: The Mind is Flat   d: Shakespeare

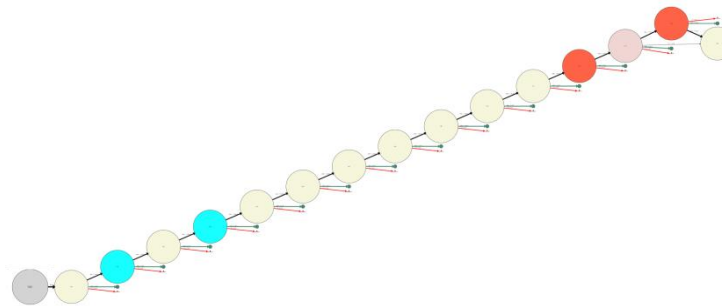

**Fig.5.** Completers learning path, first week (Shakespeare) fish eye view

Then, we chart the percentage of learning transition from different *themes* for the two learner groups (Figure 6). We compared dropout after their last accessed activity: video, discussion, quiz and article for the four courses. The figure illustrates that learners are more likely to drop out after articles and videos. Interestingly, participants drop out the least after discussion. The attraction of discussion has been confirmed by our statistical analysis across courses. The reason may be that participants feel encouraged to share their knowledge and can gain support [26]. In the Literature course, Shakespeare, participants were more likely to drop out after the assignment. The reason may be the difficulty of the creative writing at the final week that learners are required to write their film, book, ballet or musical; this is useful feedback for the course designers to change assignments potentially. The figures also suggest that the dropout patterns, according to themes are similar across runs.

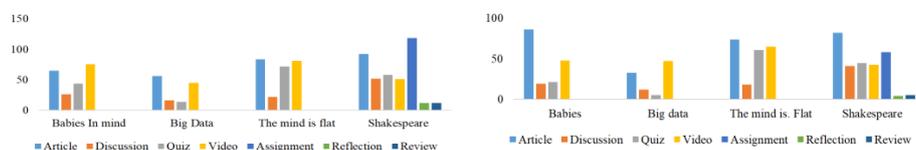

**Fig. 6** Number of dropout/topic：a) first run          b) second run

Furthermore, we predict the early dropout (dropout in first 10% of the course's duration) of the four courses based on time spend on each activity via two machine learning algorithms. The higher indicator of dropout per video shown in Figure 6, the higher prediction accuracy demonstrated in Table 3, which indicates video theme is more predicable at an early stage of learning. For instance, the numbers of dropout per video in the Mind is Flat, Babies in Mind, Big Data and Shakespeare four courses in second run are: 65, 48, 47, 42 and the accuracy also follows this order.

**Table 3.** Early Prediction (in first ten percentage of course) of Dropout

|  | **The mind is flat** | **Babies in Mind** | **Big Data** | **Shakespeare** |
|---|---|---|---|---|
| **XG Boosting** | 89.3% | 86.5% | 83.6% | 83.2% |
| **Gradient Boosting** | 89.3% | 86.5% | 83.7% | 83.5% |

To explore the relationship between learning themes at a macro, i.e. bird-eye level, we visualise the catch-up transition behaviour for dropout learners (Figure 7; as most completers transit from themes as designed).

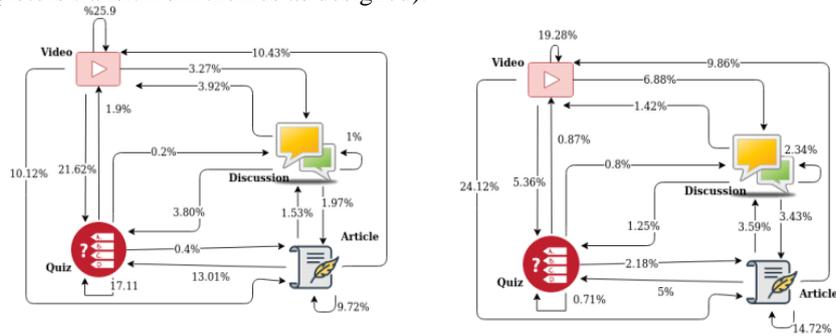

**Fig.7.** Babies in Mind (left) & Big Data (right): catch-up themes transition, dropout learners

The most significant component, 25.9% and 24.12% of the participants continue from videos to videos and switch from videos to articles in the Babies in Mind and Big Data courses, respectively. The transition from videos to articles is 24.12% in the Big Data course, i.e., nearly one-quarter of the dropout learners lose interests after reading papers. Therefore, course designers should consider replacing some readings with interactive activities, to improve engagement. Interestingly, 17.1% dropouts transfer from quizzes to quizzes, which could be inefficient attempts to finish quickly. Such behaviour can inform instructors to intervene early.

## 5    Conclusion and Limitations

The paper visualises and compares different learning paths of completers and non-completers across four MOOCs, and explores from which learning theme learners tend to drop out. We have shown how different granularity visualisation (*fish eye*, *bird eye*) allows both researchers, and potentially teachers, to understand where issues occur and where patterns emerge, backed up by statistical analysis. Specifically, we have shown that course completers are more likely to learn *linearly*, while the dropout learners are more likely to jump forward to a later activity, which we dubbed here as "*catch-up*" learning pattern. Additionally, we find that video learning pattern is a powerful indicator for early prediction of dropout for learners, validated by two machine learning models, XG Boosting and Gradient Boosting. Our research has shown the benefits of *theme*-based visualisation, based on which we have identified that non-completers are more likely to watch videos and skip quizzes, shown by their transition from other themes, to videos. Moreover, we show how this type of analysis can generate fine-grained ideas for instructors and course designers; e.g., to improve retention, instructors (and online course designers) should introduce more discussion support mechanisms.

Limitations include clickstream analysis not entirely reflecting the true engagement of learners with learning activities (as learners may, e.g., click the complete button too soon). Additionally, we only consider the first access, due to our data limitations, which may underestimate jumping patterns of dropout learners. Other learning features could be considered, e.g., length of activity accessed.